\documentclass[aps,pra,reprint,superscriptaddress,nofootinbib,floatfix]{revtex4-2}

\usepackage{amsmath,amssymb,bm}
\usepackage{booktabs}
\usepackage{graphicx}
\usepackage{microtype}
\usepackage{xcolor}
\usepackage[colorlinks=true,linkcolor=blue,citecolor=blue,urlcolor=blue]{hyperref}
\hypersetup{
  pdftitle={Adaptive operator-generated subspaces for effective many-body Hamiltonians},
  pdfauthor={Ginanjar Utama and Hermawan Kresno Dipojono},
  pdfsubject={A reproducible effective-Hamiltonian to many-body observables workflow}
}

\newcommand{\Tr}{\operatorname{Tr}}
\newcommand{\Cl}{\mathrm{Cl}}

\newcommand{\ket}[1]{\lvert #1\rangle}
\newcommand{\bra}[1]{\langle #1\rvert}
\newcommand{\avg}[1]{\langle #1\rangle}
\newcommand{\supp}{\operatorname{supp}}
\newcommand{\rank}{\operatorname{rank}}

\begin{document}

\title{Adaptive operator-generated subspaces for effective many-body Hamiltonians}

\author{Ginanjar Utama}
\email{ginanjar.utama@gmail.com}
\affiliation{Department of Engineering Physics, Institut Teknologi Bandung, Bandung, Indonesia}

\author{Hermawan Kresno Dipojono}
\email{dipojono@itb.ac.id}
\affiliation{Department of Engineering Physics, Institut Teknologi Bandung, Bandung, Indonesia}

\date{August 1, 2026}

\begin{abstract}
Electronic-structure and embedding calculations naturally terminate in an
effective many-body Hamiltonian, whereas quantum eigensolver demonstrations
often begin from a hand-built qubit model and end at a ground-state energy.
We connect these boundaries with the Adaptive Clifford-Algebra Subspace
Eigensolver (A-CASE), a single-reference, operator-generated Rayleigh--Ritz
method.  For generators $A_i$, the overlap, Hamiltonian, and observable
matrices are reconstructed as expectations of
$A_i^\dagger A_j$, $A_i^\dagger H A_j$, and
$A_i^\dagger Q A_j$ on one reference state.  A matrix-element bank reuses the
global Pauli-word universe, while adaptive growth ranks a candidate by the
energy lowering of its overlap-aware $2\times2$ generalized eigenproblem and
rejects sector leakage or near-linear dependence.  A strict, dependency-light
FCIDUMP adapter supplies a reproducible active-space boundary.  On linear
H$_4$ in STO-3G with a CAS(4e,4o) FCIDUMP, the mapped $(N=4,S_z=0)$ sector
energy agrees with an external determinant-FCI result to
$3.1\times10^{-15}$ Ha.  Eight adaptive additions give a nine-dimensional
subspace with a $3.019$ mHa error; the complete 27-dimensional
singles/doubles subspace reaches $0.766$ mHa.  Holding that budget fixed and
replacing the determinant reference by a two-operator ADAPT-VQE state, which
is itself $27.091$ mHa, moves the nine-vector error to $0.342$ mHa---inside
chemical accuracy.  This is a fixed-A-CASE-budget statement for this H$_4$
pool, not a matched total-cost claim: the hybrid additionally pays for two
state-preparation rotors, pool-gradient evaluations, and parameter
optimization.  Under a fully matched contract---same input, sector, reference,
pool, budget, and stopping rule---the operator route needs one state
preparation against ADAPT-VQE's ninety, at roughly an order of magnitude more
measured words, and the same retained subspace costs $7371$ or $2240$ words
depending only on whether it is generated by determinant excitations or by the
Pauli words they decompose into, so a measurement width quoted without its
generator resolution is not comparable across methods.
An exact ADAPT-GCIM comparator on the same local excitation pool gives
$13.364$ mHa at the nearest size match ($k=4,M=8$) and $10.674$ mHa at the
iteration match ($k=8,M=16$); its 36/28 and 136/120 Hamiltonian/overlap
state-pair counts are reported separately from the single-reference word
universe.
A broader ladder is deliberately mixed: fixed Krylov bases are much
more accurate on the molecular rungs and, once their word universe is computed
through an identity that avoids the quadratic element route, usually narrower
as well, though they carry overlap condition numbers of $10^8$--$10^{10}$;
A-CASE closes the $2\times2$ Hubbard sector only after compound/configuration
generators are admitted, and the same extension provides no improvement on
$2\times3$.
For response, one qubit-wise-commuting measurement cache reconstructs
$(S,H,Q)$, and a grouped bootstrap reruns thresholding, diagonalization, root
matching, spectral weights, susceptibility, and broadening.  The resulting
percentile bands are explicitly heuristic, pointwise, and conditional on
replicas that preserve rank and root identity; they are not finite-sample
confidence certificates.  Holding system, observable, words, groups, shots,
basis size, and seeds fixed and varying only the generator family, acceptance
falls from $200/200$ at $\kappa_S\approx1$ to $139/200$ at
$\kappa_S=9.3\times10^{3}$, with every rejection a moved thresholded rank.
The work establishes an executable and
independently checked path from an interchange Hamiltonian to energies,
correlations, and response.  It does not establish materials accuracy,
favorable scaling, or a quantum advantage.
\end{abstract}

\maketitle

\section{Introduction}
\label{sec:introduction}

The difficult part of a materials calculation is not represented by a single
algorithm.  Density-functional theory, Wannierization, constrained screening,
and embedding each remove degrees of freedom or redefine interactions before
a correlated active-space problem is solved
\cite{marzari2012wannier,pizzi2020wannier90,aryasetiawan2004crpa,georges1996dmft}.
The output of that upstream chain is an effective many-body Hamiltonian, not a
quantum circuit.  Conversely, variational and subspace quantum algorithms are
often assessed on a qubit Hamiltonian already in memory
\cite{peruzzo2014vqe,mcclean2017qse,huggins2020nonorthogonal}.  A useful
downstream solver must therefore make its input contract as explicit as its
eigensolver: orbital order, electron number, spin sector, energy units,
one- and two-body conventions, provenance, and the reference determinant all
belong to the scientific result.

Quantum subspace diagonalization (QSD) is attractive at this boundary because
it replaces a deep optimized circuit by matrix elements in a small,
nonorthogonal subspace.  Quantum subspace expansion (QSE), nonorthogonal VQE,
and quantum Krylov methods instantiate this idea with different basis
families \cite{mcclean2017qse,huggins2020nonorthogonal,stair2020krylov}.
Iterative and residual-driven variants aim to improve compactness
\cite{tkachenko2024davidson,oleary2025partitioned}, while QSCI selects a
determinant subspace from sampled configurations
\cite{nakagawa2024adaptqsci}.  IQAE iteratively enlarges a Pauli/Hamiltonian
moment space \cite{bharti2021iqae}; Q-SENSE exchanges circuit depth for
seniority-structured matrix elements \cite{patel2026qsense}; and AS-SQD uses a
finite-shot, perturbative acquisition rule for sampled determinant spaces
\cite{miura2026assqd}.  Most closely, ADAPT-GCIM selects UCC/Givens generating
functions adaptively, solves a nonorthogonal generalized eigenproblem, and can
be composed with ADAPT-VQE \cite{zheng2024adaptgcim}.  Thus neither adaptive
subspace growth nor replacing variational optimization by a generalized
eigenproblem is new in this work.  Two obstacles recur across these methods.
First, a compact basis in vector count can still be expensive in measurement
width.  Second, the generalized eigenproblem
\begin{equation}
    H_{\rm sub}c = E S c
    \label{eq:gep-intro}
\end{equation}
can become ill-conditioned, so small measurement errors are amplified
\cite{epperly2022theory,lee2024sampling,zhang2024measurement}.  Reporting only
$M$, the number of basis vectors, conceals both costs.

This work studies an adaptive operator-generated construction, A-CASE:
\emph{Adaptive Clifford-Algebra Subspace Eigensolver}.  The method uses one
reference density operator $\rho$ and virtual basis vectors
$A_i\ket{\psi}$; it does not prepare a different quantum state for every
$i$.  Every matrix element is reduced to Pauli-word expectations on $\rho$,
and the same cached expectations support energy, projected observables, and
response.  The complex Clifford algebra
$\Cl(2n,\mathbb C)\cong M(2^n,\mathbb C)$ supplies the implementation's typed
operator representation, with Jordan--Wigner strings mapping fermionic modes
to local qubit Paulis \cite{utama2026operatorcentric}.  This is a
representation choice, not extra expressivity and not a speedup claim.

The contribution is fourfold.  First, we specify a strict FCIDUMP-to-qubit
boundary and independently check its integral convention.  Second, we make
adaptive growth overlap-aware and resource-aware: candidate selection,
conditioning, symmetry leakage, word-universe growth, and retained rank are
recorded together.  Third, we connect the same projected space to
correlations and Lehmann response without materializing the Ritz state.
Fourth, we propagate finite-shot variability through the \emph{whole}
nonlinear response pipeline by grouped resampling, while refusing to call the
result a confidence certificate.

Table~\ref{tab:relation} states the methodological boundary against the
closest approaches.  In particular, A-CASE is distinguished from ADAPT-GCIM
by its fixed-reference Pauli-expectation bank and its overlap-aware local
energy-lowering score, not by adaptivity or the generalized eigenproblem.
The matched H$_4$ ladder includes an exact implementation of ADAPT-GCIM's
published selector and generating-function rule: unused anti-Hermitian
excitations are ranked on the cumulative surrogate state, every selected
unitary uses the fixed angle $\theta=\pi/4$, and the basis grows as $M=2k$
without parameter optimization.  This is an algorithm-level comparison on the
same local pool, not a reproduction of the published molecular curves or a
numerical superiority claim over that method in its original setting.

\begin{table*}[t]
\caption{Relation to the closest adaptive and operator-generated subspace
methods.  ``Prepared objects'' identifies what must change across basis
directions; ``selection'' identifies the adaptive acquisition rule.}
\label{tab:relation}
\begin{ruledtabular}
\footnotesize
\begin{tabular}{@{}p{0.17\textwidth}p{0.17\textwidth}p{0.17\textwidth}p{0.17\textwidth}p{0.17\textwidth}@{}}
Method & Basis object & Selection & Prepared objects & Primary distinction here \\
\colrule
IQAE \cite{bharti2021iqae} & Hamiltonian/Pauli moments & Iterative moment hierarchy & One reference & No response or word-cost selector \\
ADAPT-GCIM \cite{zheng2024adaptgcim} & UCC/Givens generating functions & Surrogate-state gradient & Generating-function circuits & Closest adaptive GEP architecture \\
Q-SENSE \cite{patel2026qsense} & Seniority-transformed states & Symmetry hierarchy & Short unitary circuits & Orthogonality from seniority sectors \\
AS-SQD \cite{miura2026assqd} & Sampled determinants & Perturbative energy acquisition & Bitstring samples & Finite-shot determinant expansion \\
A-CASE (this work) & Virtual operators on one reference & Overlap-aware local pencil with word cost & One fixed reference & Shared bank for energy, observables, response \\
\end{tabular}
\end{ruledtabular}
\end{table*}

\begin{figure*}[t]
    \includegraphics[width=\textwidth]{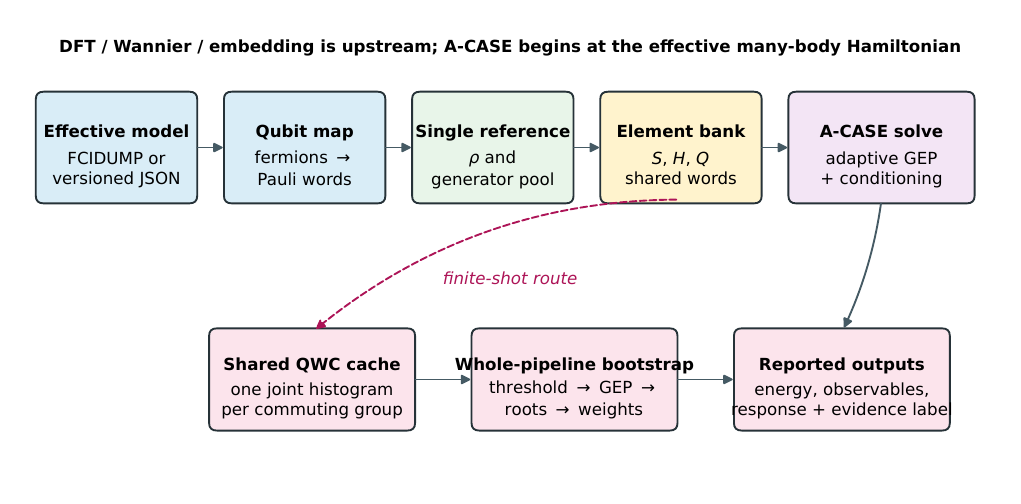}
    \caption{Computational contract.  The exact route builds one global
    matrix-element bank and solves an adaptively grown generalized eigenproblem.
    The finite-shot route measures the union word set in QWC groups and
    resamples each group's joint histogram through every nonlinear downstream
    operation.  Upstream electronic-structure and embedding stages are outside
    the claimed scope.}
    \label{fig:pipeline}
\end{figure*}

Figure~\ref{fig:pipeline} summarizes the boundary.  A-CASE begins at an
effective Hamiltonian.  It is not a DFT, Wannier, cRPA, or DMFT implementation,
and the synthetic dimer below is not represented as the result of any of those
upstream calculations.

\section{Operator-generated adaptive subspaces}
\label{sec:method}

\subsection{Single-reference matrix elements}
\label{sec:elements}

Let $H=\sum_w h_w P_w$ be an $n$-qubit Hamiltonian and
$\rho=\ket{\psi}\!\bra{\psi}$ a normalized reference density operator.
For an ordered generator family $\{A_i\}_{i=0}^{M-1}$ with $A_0=I$, define
\begin{align}
    S_{ij} &= \Tr(\rho A_i^\dagger A_j), \label{eq:s}\\
    H_{ij} &= \Tr(\rho A_i^\dagger H A_j), \label{eq:h}\\
    Q_{ij} &= \Tr(\rho A_i^\dagger Q A_j). \label{eq:q}
\end{align}
The retained Ritz vectors solve Eq.~\eqref{eq:gep-intro} after the overlap
spectrum is thresholded.  Their observable matrix elements are
\begin{equation}
    \avg{Q}_{kl}
    =\frac{c_k^\dagger Q_{\rm sub}c_l}
    {\sqrt{(c_k^\dagger S c_k)(c_l^\dagger S c_l)}}.
    \label{eq:projected-q}
\end{equation}
No reconstructed $2^n$-component Ritz state is needed.

Each operator in Eqs.~\eqref{eq:s}--\eqref{eq:q} is expanded in the Hermitian
Pauli basis.  Although $A_i^\dagger H A_j$ need not be Hermitian, its
coefficients may be complex while each Pauli expectation is real.  Thus every
entry is a linear reconstruction from expectations measured on the
\emph{same} $\rho$; neither an ancilla-based Hadamard test nor a separately
prepared $A_j\ket{\psi}$ is required.  This observation does not remove the
cost: it moves the relevant cost boundary to the number and grouping of
distinct Pauli words.

\subsection{Matrix-element bank and resource boundary}
\label{sec:bank}

For a basis index set $\mathcal B$, the measurement universe is
\begin{align}
 \mathcal W(\mathcal B)
 &=
 \bigcup_{i,j\in\mathcal B}\supp(A_i^\dagger A_j)
 \nonumber\\
 &\quad\cup
 \bigcup_{i,j\in\mathcal B}\supp(A_i^\dagger H A_j).
 \label{eq:word-universe}
\end{align}
Its width is $W=|\mathcal W|$.
The matrix-element bank caches the operator-valued entries, reuses Hermitian
pairs, and records whether an accepted generator reuses an existing word or
introduces a new one.  Projected observables extend the same bank by
$\supp(A_i^\dagger Q A_j)$.  The relevant resource ledger is therefore
\begin{equation}
 (M,\;\rank S,\;\kappa_S,\;W,\;S_H,\;S_A,\;\ell_N,\;\ell_{S_z}),
 \label{eq:ledger}
\end{equation}
where $S_H$ is the largest support of a cached Hamiltonian element, $S_A$ the
largest generator support, and $\ell_N,\ell_{S_z}$ are relative commutator
norms with particle number and spin projection.  A method is not compact merely
because $M$ is small.

\subsection{Pricing the Krylov baseline}
\label{sec:krylov-width}

A ledger that reports $W$ for one method and leaves it blank for the
comparator cannot support a claim about measurable support.  The obstruction is
that building the element operators costs $O(|A_i|\,|H|\,|A_j|)$ per pair, and
deep Hamiltonian powers carry thousands of words each.  For the Krylov family
that cost is avoidable.  With $A_0=I$, $A_k=H^k$, and $H$ Hermitian,
\begin{equation}
 A_i^\dagger A_j = H^{i+j},\qquad
 A_i^\dagger H A_j = H^{i+j+1},
 \label{eq:krylov-collapse}
\end{equation}
so the union in Eq.~\eqref{eq:word-universe} reduces from $O(M^2)$ distinct
element operators to the $2m+2$ powers $H^0,\dots,H^{2m+1}$.  This is linear in
$M$ and runs in seconds.  On the rungs small enough for the direct enumeration
to finish, the two agree exactly, including against the ladder's own tracked
counts of $24$ and $64$ words for H$_2$ and LiH.

One caveat is quantitative rather than structural.  Repeated multiplication
accumulates round-off, and after seventeen products thousands of words carry
coefficients near $10^{-11}$ that are absent from the exact operator.  The
A-CASE universe is insensitive to this---every one of its $7371$ words on
H$_4$ survives a $10^{-8}$ cut---while the raw Krylov count nearly doubles
because of it.  The untruncated count is not even reproducible: two runs of
the same rung differ by a few words, because the upstream self-consistent
field settles on orbital coefficients differing in the last bits and a
Hamiltonian perturbed at $10^{-16}$ has a different round-off-level support at
the seventeenth power.  A cutoff is an approximation, so it cannot be paired
with the untruncated Krylov energy without a further check.  Every cutoff in
the committed sweep is therefore applied independently to the same unpruned
powers, never fed recursively into the next multiplication; the support sets
are consequently nested.  At each cutoff we rebuild the complete Hankel
overlap and Hamiltonian pencils and resolve the GEP.  The reported $10^{-8}$
count is admitted only when effective rank agrees and the normalized pencil
entries, ground energy, and condition number meet the declared tolerances.
On every quoted rung those differences are zero at the stored
double-precision resolution.  Thus Table~\ref{tab:ladder} pairs the reported
Krylov energy and conditioning only with a support count certified to describe
the same numerical pencil.

\subsection{Adaptive growth}
\label{sec:growth}

Suppose the current normalized Ritz pair is $(E,\ket{\Psi})$ and a candidate
direction is $\ket{\chi_a}=A_a\ket{\psi}$.  A-CASE evaluates the generalized
$2\times2$ pencil
\begin{equation}
 \begin{pmatrix}E&h_a\\h_a^*&h_{aa}\end{pmatrix}v
 =\lambda
 \begin{pmatrix}1&s_a\\s_a^*&s_{aa}\end{pmatrix}v,
 \label{eq:two-by-two}
\end{equation}
with $s_a=\langle\Psi|\chi_a\rangle$,
$h_a=\langle\Psi|H|\chi_a\rangle$,
$s_{aa}=\langle\chi_a|\chi_a\rangle$, and
$h_{aa}=\langle\chi_a|H|\chi_a\rangle$.  The score is the predicted lowering
$\Delta_a=\max(0,E-\lambda_{\min})$, optionally divided by a penalty for newly
introduced words.  Keeping the overlap block is essential: replacing it by
the identity would give a nonzero score to a rescaled duplicate of the
current Ritz state.

Before scoring, the component outside the retained subspace is measured by
\begin{equation}
 \eta_a =
 1-\frac{\|{\cal P}_{\mathcal B}\chi_a\|^2}{\|\chi_a\|^2}.
 \label{eq:orthogonal-fraction}
\end{equation}
Candidates below an orthogonality floor, annihilating the reference, or
exceeding the declared sector-leakage tolerance are rejected.  The accepted
candidate with the largest deterministic score appends one row and column to
the cached pencil.  The exact Ritz energy is variational and non-increasing
under nested growth; that statement does not survive arbitrary finite-shot
perturbations of $(S,H)$.

The candidate hierarchy used here contains identity, symmetry-preserving
determinant excitations, commutator-response directions, selected Hamiltonian
powers, compound products, and explicitly reached competing-order
configurations.  These are Pauli-sum operators in a Clifford-algebra
representation; they are not quantum-information Clifford transformations.
The hierarchy is intentionally heterogeneous because no single family spans
all of the molecular, Hubbard, and spin-model rungs below.

\section{Interchange boundary and response}
\label{sec:boundary-response}

\subsection{Restricted FCIDUMP adapter}
\label{sec:fcidump}

FCIDUMP is a compact interchange record for active-space one- and two-electron
integrals \cite{knowles1989fcidump}.  For a real restricted record, the parser
constructs the chemist-notation Hamiltonian directly,
\begin{align}
 H={}&E_{\rm core}
 +\sum_{pq\sigma}h_{pq}a^\dagger_{p\sigma}a_{q\sigma}\nonumber\\
 &+\frac12\sum_{pqrs}\sum_{\sigma\tau}(pq|rs)
 a^\dagger_{p\sigma}a^\dagger_{r\tau}a_{s\tau}a_{q\sigma}.
 \label{eq:fcidump-h}
\end{align}
The implementation restores packed integral symmetries, accepts Fortran
\texttt{D} exponents, validates \texttt{NORB}, \texttt{NELEC},
\texttt{MS2}, sentinels, duplicates, finiteness, and electron--spin parity,
then maps an interleaved spin-orbital ordering through Jordan--Wigner.
Unrestricted \texttt{IUHF=1} records are rejected rather than interpreted by
guess.  A SHA-256 digest binds every benchmark result to the exact input
gauge.  An independent OpenFermion reconstruction
\cite{mcclean2020openfermion} agrees over 185 Pauli terms to a maximum
coefficient difference of $8.74\times10^{-16}$.

\subsection{Projected Lehmann response}
\label{sec:response}

For a Hermitian perturbation $Q$, the projected positive-frequency lines are
\begin{equation}
 \omega_f=E_f-E_0,\qquad
 w_f=|\langle\Psi_f|Q|\Psi_0\rangle|^2,
 \label{eq:lehmann}
\end{equation}
and the nondegenerate zero-temperature static susceptibility is
\begin{equation}
 \chi(0)=2\sum_{f>0}\frac{w_f}{\omega_f}.
 \label{eq:chi}
\end{equation}
A Lorentzian of width $\eta$ converts the discrete lines into
\begin{equation}
 S_\eta(\omega)=\sum_f
 \frac{w_f\eta}{\pi[(\omega-\omega_f)^2+\eta^2]}.
 \label{eq:broadening}
\end{equation}
These quantities are exact only for the chosen projected subspace.  Response
completeness is a separate requirement from ground-state accuracy:
$Q\ket{\Psi_0}$ must be represented by the retained roots
\cite{colless2018spectra,umeano2025response}.

\subsection{Finite-shot nonlinear uncertainty}
\label{sec:bootstrap}

One qubit-wise-commuting (QWC) partition
\cite{verteletskyi2020grouping} is built for the union of the words required
by $S$, $H$, and $Q_{\rm sub}$.  Each group contributes a joint outcome
histogram, retaining the covariance between compatible Pauli words read from
the same shots.  The finite-shot response is not a fixed linear functional:
\begin{align}
 \{\text{histograms}\}
 &\longmapsto(S,H,Q)
 \longmapsto\text{thresholded GEP}
 \nonumber\\
 &\longmapsto(\omega_f,w_f,\chi,S_\eta).
 \label{eq:nonlinear-map}
\end{align}
Linear word-mean intervals cannot be substituted through this map and called
spectral confidence intervals.

We use a grouped nonparametric bootstrap \cite{efron1979bootstrap}.  Each
replica draws a multinomial histogram within every fixed QWC group and reruns
all operations in Eq.~\eqref{eq:nonlinear-map}.  Ordered-root matching is used
only when the point estimate and the replica have isolated adjacent roots.
Replicas with a changed retained rank, a root collision, or a solver failure
are counted and excluded.  Consequently, the percentile intervals are
\emph{conditional on the surviving replicas}.  That exclusion is not neutral:
the failed draws are precisely the draws that can widen a spectrum in an
ill-conditioned regime.  We therefore report the acceptance rate beside every
interval.

The evidence label is always \texttt{heuristic}; \texttt{certified=False} is
part of the result contract.  A finite-sample certificate would require a
confidence set for the correlated matrix pencil and observable together with
a valid root-identification argument.  Pointwise intervals for
$S_\eta(\omega)$ are not a simultaneous spectral band.

\section{Benchmark design}
\label{sec:benchmarks}

\subsection{A closed-form integration oracle}
\label{sec:dimer}

The smallest end-to-end model is a half-filled two-site Hubbard dimer with
$t=1$ eV and $U=4$ eV.  A versioned JSON record plays the role of a synthetic
two-Wannier-site effective Hamiltonian.  Its $(N=2,S_z=0)$ sector has dimension
four, and the correlated singlet is analytically solvable.  Table~\ref{tab:dimer}
checks every reported energy, correlation, and response quantity against a
closed form that shares no projected-observable code.

\begin{table*}[t]
\caption{Two-site effective-Hamiltonian oracle.  The response basis has
$M=4$, equal to the sector dimension; it is full configuration interaction in
that sector and is not a compactness result.}
\label{tab:dimer}
\begin{ruledtabular}
\begin{tabular}{lcc}
Quantity & A-CASE result & independent check \\
\colrule
Ground energy $E_0$ & -0.828427125 eV & $(U-\sqrt{U^2+16t^2})/2$ \\
Double occupancy per site $d$ & 0.073223305 & $(\partial E_0/\partial U)/2$ \\
$\langle\mathbf S_0\!\cdot\!\mathbf S_1\rangle$ & -0.640165043 & $-\frac34(1-2d)$ \\
$\langle S^2\rangle$ & $<10^{-12}$ & singlet invariant \\
Staggered-spin gap $\omega$ & 0.828427125 eV & $-E_0$ \\
Staggered-spin weight $w$ & 0.853553391 & $1-2d$ \\
Static susceptibility $\chi(0)$ & 2.060660172 eV$^{-1}$ & $2w/\omega$ \\

\end{tabular}
\end{ruledtabular}
\end{table*}

\subsection{Frozen \texorpdfstring{H$_4$}{H4} active space}
\label{sec:h4-benchmark}

The interchange benchmark is linear H$_4$ with separations of $0.9$ \AA,
STO-3G, and CAS(4e,4o): four electrons in four spatial orbitals, mapped to
eight qubits.  The frozen FCIDUMP was generated with PySCF 2.14
\cite{sun2018pyscf} using an RHF convergence tolerance of $10^{-12}$.
The file digest, geometry, basis, software version, RHF energy, and an external
determinant-FCI energy are committed.  After checkout, parsing, mapping,
sector diagonalization, and A-CASE require only NumPy.

\begin{table*}[t]
\caption{Frozen H$_4$ FCIDUMP benchmark.  The adaptive budget is eight
additions.  ``Complete singles/doubles'' is a fixed determinant-generated
subspace and is shown as a reproducible accuracy rung, not as adaptive
selection.}
\label{tab:h4}
\begin{ruledtabular}
\begin{tabular}{lccc}
Method & Energy (Ha) & Error & $M$ or sector dimension \\
\colrule
External determinant FCI & -2.180316614323862 & --- & --- \\
Mapped sector oracle & -2.180316614323859 & $3.11\times10^{-15}$ Ha & 36 \\
Adaptive A-CASE & -2.177297832980223 & 3.018781 mHa & 9 \\
Complete singles/doubles & -2.179550752062548 & 0.765862 mHa & 27 \\

\end{tabular}
\end{ruledtabular}
\end{table*}

\subsection{The reference state as a variable}
\label{sec:warm-start}

The adaptive rung above grows its subspace around a single Hartree--Fock
determinant, so every correlation effect must be paid for out of the eight
additions.  That is a choice, not a requirement: any state the pipeline can
prepare may serve as $\rho$.  We therefore repeat the H$_4$ benchmark with the
budget, candidate pool, and frozen FCIDUMP held fixed and the reference
replaced by an ADAPT-VQE state \cite{grimsley2019adapt} of $k$ operators,
obtained from a qubit-ADAPT pool built from the odd-$Y$ words of the same
determinant excitations.  Table~\ref{tab:warm} reports the ADAPT stage's own
energy beside the A-CASE result, because the two must not be conflated.
It also reports the pool-gradient and optimizer evaluations and the number of
ADAPT rotors that must be executed before every A-CASE measurement.  These are
exact-simulation algorithmic counts.  Because selection and optimization are
noiseless here, assigning them zero shots would be a simulation convention,
not a physical-cost estimate; no end-to-end hardware shot comparison is made.

\begin{table*}[t]
\caption{Hybrid resource ledger.  The frozen FCIDUMP, A-CASE candidate pool,
and nine-vector A-CASE budget are shared, but the ADAPT preparation is
additional work.  ``Grad.'' is the total number of active-pool gradients
evaluated; ``opt.'' is the optimizer's joint energy/gradient evaluations;
``rot.'' is the number of state-preparation rotors.  Both energy-error columns
are in mHa.  These are exact-simulation counts, not a physical shot estimate.}
\label{tab:warm}
\begin{ruledtabular}
\footnotesize
\begin{tabular}{lcccccccc}
Reference $\rho$ & Grad. & Opt. & Rot. & ADAPT error & $M$ & A-CASE error & $\kappa_S$ & $W$ \\
\colrule
Hartree--Fock determinant & 0 & 0 & 0 & --- & 9 & 3.019 & 1 & 7371 \\
ADAPT-VQE state, $k=2$ & 319 & 15 & 2 & 27.091 & 9 & 0.342 & 1.02 & 7510 \\
ADAPT-VQE state, $k=4$ & 634 & 36 & 4 & 13.779 & 9 & 0.612 & 1.05 & 7519 \\
ADAPT-VQE state, $k=6$ & 945 & 57 & 6 & 6.791 & 9 & 0.768 & 1.06 & 7510 \\

\end{tabular}
\end{ruledtabular}
\end{table*}

\subsection{One contract, explicit currencies}
\label{sec:matched}

Table~\ref{tab:warm} varies $\rho$ at a fixed A-CASE budget, which prices the
warm start against cold A-CASE but not against the alternatives.  We therefore
fix one contract and run every arm inside it: the same frozen FCIDUMP and
digest, the same $(N=4,S_z=0)$ sector, the same Hartree--Fock reference, the
same eight-addition budget, and exact arithmetic throughout.  Every arm runs
its budget out, with no early exit.  The pool is the same operator content for
every arm: 26 symmetry-preserving determinant excitations, which decompose into
160 odd-$Y$ Pauli words.  ADAPT-GCIM necessarily needs two rows: its published
basis rule gives $M=2k$, so $k=4$ gives the nearest basis-size match ($M=8$)
to the nine-vector subspaces, while $k=8$ matches the number of adaptive
iterations and gives $M=16$.

Granularity is the one thing the contract cannot fix.  ADAPT-VQE consumes
single words; A-CASE's default candidates are whole excitations.  Rather than
declare one canonical, we run A-CASE at both resolutions.  ADAPT-GCIM uses the
whole-excitation resolution.  Costs are reported as separate currencies rather
than summed, because a state preparation, a Pauli word, and an off-diagonal
state-pair measurement are different machines' bottlenecks and an exchange
rate between them would be an assumption, not a measurement.

\begin{table*}[t]
\caption{Matched H$_4$ contract.  ``Prep.'' counts distinct states the arm must
prepare; circuit executions are that times the QWC group count times shots per
group, and the shot factor is outside this exact-arithmetic record.  ``Sel.''
is candidate scorings summed over steps and ``Sel.\ $W$'' the words needed to
perform them---for A-CASE the whole element cache, including rows for
candidates later rejected, which is strictly larger than the retained $W$ the
ledger of Eq.~\eqref{eq:ledger} reports.  ``$H/S$ pairs'' gives unique
upper-triangle Hamiltonian pairs/off-diagonal overlap pairs for ADAPT-GCIM;
these transition measurements do not form a single-reference $W$.
``---'' marks a column an arm does not have and ``n/t'' a word count that is
not the applicable measurement primitive.  ``Rot.'' is the deepest prepared
state; the record also stores the sum over all distinct basis circuits.}
\label{tab:matched}
\scriptsize
\begin{ruledtabular}
\begin{tabular}{lccccccccc}
Arm & $M$ & Error (mHa) & $\kappa_S$ & Prep. & Rot. & Sel. & Sel.\ $W$ & $W$ & $H/S$ pairs \\
\colrule
QSE & 9 & 41.653 & 1 & 1 & 0 & --- & --- & 4,226 & --- \\
Krylov & 9 & $1.05\times10^{-5}$ & $6.6\times10^{10}$ & 1 & 0 & --- & --- & 4,224\footnotemark[1] & --- \\
generator coordinate & 9 & 49.576 & 1 & 1 & 0 & --- & --- & 13,646 & --- \\
ADAPT-GCIM (4 iter., M=8) & 8 & 13.364 & $3.3\times10^{1}$ & 8 & 4 & 98 & 2,424 & n/t & 36/28 \\
ADAPT-GCIM (8 iter., M=16) & 16 & 10.674 & $1.1\times10^{3}$ & 16 & 8 & 180 & 2,424 & n/t & 136/120 \\
ADAPT-VQE & --- & 2.375 & --- & 90 & 8 & 1,252 & 2,424 & 185 & --- \\
A-CASE (determinant) & 9 & 3.019 & 1 & 1 & 0 & 180 & 15,783 & 7,371 & --- \\
A-CASE (word, leakage rejected) & 1 & 56.057 & 1 & 1 & 0 & --- & 185 & 185 & --- \\
A-CASE (word) & 9 & 3.019 & 1 & 1 & 0 & 1,252 & 14,401 & 2,240 & --- \\
A-CASE (determinant, ADAPT warm start) & 9 & 0.342 & 1.02 & 18 & 2 & 499 & 15,803 & 7,510 & --- \\

\end{tabular}
\footnotetext[1]{From Sec.~\ref{sec:krylov-width}; the tracked element route
does not reach the deep Krylov powers.}
\end{ruledtabular}
\end{table*}

Four readings follow, and only the first is the one the method was designed
to produce.

\emph{The preparation asymmetry is real and large.}  Every subspace arm needs
one prepared state only when its basis is virtual on a fixed reference:
A-CASE and the fixed operator bases do, whereas ADAPT-GCIM prepares 8 or 16
distinct generating-function states and ADAPT-VQE needs 90 state
preparations, one per selection step and optimizer evaluation.  That is the
architectural claim of Sec.~\ref{sec:elements} in numbers.  It is bought, not
free: ADAPT-VQE reads 2424 words to score its pool and 185 to measure its final
energy, against A-CASE's 14401--15783 and 2240--7371.  The fixed-reference
operator route trades a $90\times$ reduction in state preparations for roughly
an order of magnitude more measured words.  Which side of that trade is
favorable is a hardware question this paper does not answer.

\emph{The exact ADAPT-GCIM comparison changes the baseline, not the
conclusion.}  At the nearest size match, $k=4$ gives $M=8$, effective rank 6,
$\kappa_S=32.9$, and a $13.364$ mHa error.  At the iteration match, $k=8$
gives $M=16$, effective rank 12, $\kappa_S=1.09\times10^3$, and a
$10.674$ mHa error.  Both are variationally above the sector oracle, and the
larger nested basis lowers the energy.  There is no optimizer: the selector
uses 98 and 180 candidate gradients, respectively, with the published fixed
angle.  The final pencils require 36/28 and 136/120 unique $H/S$ state pairs,
so their cost cannot be inferred from A-CASE's single-reference $W$.  These
numbers establish the exact algorithm on the matched 26-excitation local pool;
they do not reproduce ADAPT-GCIM under a larger generalized pool or convert
state pairs into a hardware shot count.

\emph{The measurement width is a property of the representation, not of the
subspace.}  Run on the determinant excitations, adaptive A-CASE needs $7371$
words; run on the 160 words those excitations decompose into, it needs $2240$.
The two arms are not merely close.  They return the same energy to
$4\times10^{-16}$ Ha, the same $M=9$ and $\kappa_S=1$, and their retained
subspaces have all nine principal angles zero---they are the same subspace,
reached by generators of different granularity, at $3.3\times$ different
measurement cost.  The ledger column introduced to stop $M$ from flattering a
method is therefore itself representation-dependent, and a $W$ quoted without
its generator resolution is not comparable across methods.  At word resolution
A-CASE needs fewer words than the Krylov arm's $4224$, which reverses the
width comparison of Sec.~\ref{sec:results-ladder} on this rung; the width
saving is paid for in selection, $1252$ scorings against $180$.  The effect is
not particular to this rung: on the $2\times2$ Hubbard plaquette the same
substitution takes $2519$ words to $413$, a factor of six, again at an
unchanged basis size and energy.

\emph{The leakage rule is stricter than the physics.}  The word pool is
rejected outright under the declared tolerance---every odd-$Y$ word has
operator leakage $\sqrt2$ out of particle number---leaving $M=1$ and the
Hartree--Fock energy.  Yet with the rule disabled the resulting Ritz vector
has sector weight $1$ to machine precision.  The test asks whether a generator
could take \emph{any} reference out of the sector, which for a determinant
reference is stronger than what is needed: a single word acting on a
determinant returns a determinant.  The conservative rule is the right default
for an arbitrary reference and is leaving a factor of three on the table here.

The exception proves the rule's purpose.  The warm-started arm is the only one
whose Ritz vector is not exactly in sector, at weight $0.999882$: its reference
is an ADAPT state assembled from those same symmetry-breaking words, and a
product of rotors does not return a determinant.  The leak is small and does
not threaten the energy, but it is real, and it is the situation the
operator-level test exists to catch.

\subsection{Validation ladder and matched baselines}
\label{sec:ladder}

The broader 72-run committed ladder covers H$_2$, LiH(2e,2o), equilibrium and
stretched H$_4$, stretched H$_2$O CAS(4e,4o), H$_2$O CAS(8e,6o), open-boundary
$2\times2$ and $2\times3$ Hubbard clusters, and a $2\times2$ Kitaev cluster.
The comparison set is the reference determinant, sector-exact diagonalization,
fixed Pauli QSE, Hamiltonian-power Krylov, a strided generator-coordinate
basis, exact ADAPT-VQE \cite{grimsley2019adapt}, and A-CASE.  Default exact
subspace arms use
$M=9$ wherever available.  Level-4 A-CASE is additionally run to $M=26$ on
the Hubbard clusters.  Chemical accuracy is defined as $1.6$ mHa only on
molecular rungs; lattice-model errors remain in the Hamiltonian's own units.

The paper uses four evidence labels:
\begin{itemize}
 \item \texttt{reference}: sector-exact energy against which errors are formed;
 \item \texttt{exact}: noiseless arithmetic on a declared subspace;
 \item \texttt{finite\_sample}: only the stated sample-split growth decision
 carries a finite-sample bound; the noisy Ritz energy is not certified or
 variational;
 \item \texttt{heuristic}: grouped-bootstrap response diagnostics.
\end{itemize}
This taxonomy prevents a confidence statement about one selection event from
silently becoming a confidence statement about the final spectrum.

\section{Results}
\label{sec:results}

\subsection{The interchange boundary is independently reproducible}
\label{sec:results-boundary}

The dimer reaches the analytic sector ground energy to
$6.7\times10^{-16}$ eV, and all six nontrivial observables in
Table~\ref{tab:dimer} agree to nine printed digits.  This establishes the
software path from a versioned effective Hamiltonian through energies,
coefficients, correlations, and response.  It does not establish materials
accuracy because the input parameters are synthetic.

For H$_4$, Table~\ref{tab:h4} separates three questions.  The mapped sector
oracle agrees with external determinant FCI to $3.1\times10^{-15}$ Ha, testing
the FCIDUMP convention and fermion-to-qubit map.  At the predeclared
nine-vector budget, adaptive A-CASE recovers most but not all of the
correlation energy and misses chemical accuracy at $3.019$ mHa.  The complete
27-dimensional singles/doubles subspace reaches $0.766$ mHa.  Because the
full $(N=4,S_z=0)$ sector has dimension 36, this is a meaningful active-space
rung, but it remains an eight-qubit calculation and is not evidence of
favorable scaling.

Table~\ref{tab:warm} shows that, for this H$_4$ instance and candidate family,
changing the reference is sufficient to cross chemical accuracy without
increasing the nine-vector A-CASE budget.  Replacing the Hartree--Fock
determinant by a
two-operator ADAPT-VQE state moves the error from $3.019$ mHa to $0.342$ mHa,
inside chemical accuracy and below the 27-dimensional complete
singles/doubles rung, while $\kappa_S$ stays near unity and $W$ grows by under
two percent.  The ADAPT state is not doing the work: on its own it is
$27.091$ mHa, an order of magnitude worse than the cold A-CASE result it
rescues.  What changes is that the eight additions no longer have to spend
themselves reaching correlation the reference could have carried.  This does
not make the hybrid free: Table~\ref{tab:warm} exposes the extra gradient,
optimization, and state-preparation work, and the present exact simulation
does not convert those counts to hardware shots.

The effect is not monotonic in ADAPT depth.  Deeper warm starts are better
states---$13.779$ mHa at $k=4$ and $6.791$ mHa at $k=6$---and yet yield worse
A-CASE results, $0.612$ and $0.768$ mHa.  A plausible reading is that the
determinant-excitation pool is matched to a determinant, so a reference further
from one is less well served by those candidates; we have not tested that, and
it identifies reference and pool as coupled choices rather than independent
ones.

\subsection{Compactness is a three-way trade}
\label{sec:results-ladder}

\begin{figure*}[t]
    \includegraphics[width=0.94\textwidth]{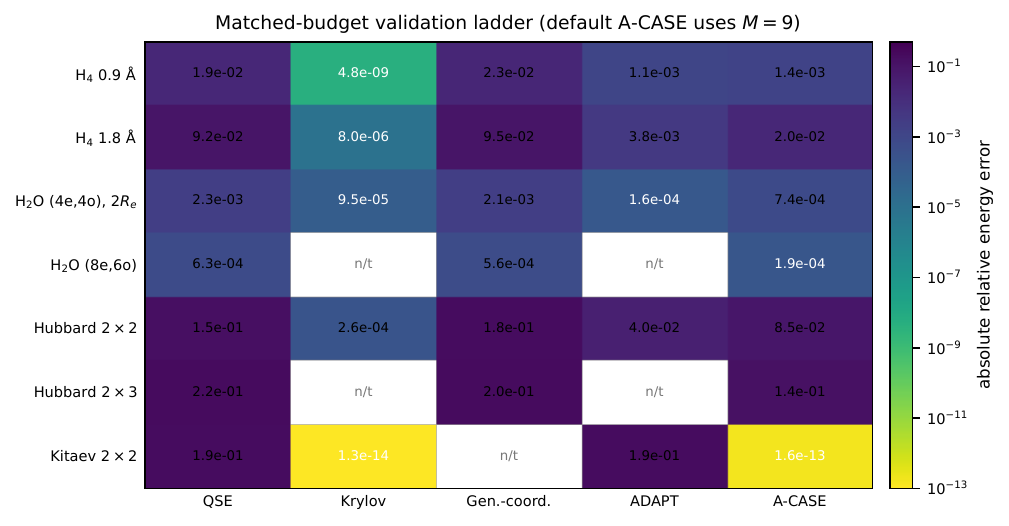}
    \caption{Absolute relative energy error for the default matched-budget
    ladder.  ``n/t'' means that the arm was not run.  The heat map compares
    energy only; Table~\ref{tab:ladder} supplies the conditioning and
    measurement-width trade.  The default A-CASE column uses $M=9$, not the
    larger level-4 Hubbard run.}
    \label{fig:ladder}
\end{figure*}

Figure~\ref{fig:ladder} shows the default matched-budget ladder in dimensionless
relative error.  Table~\ref{tab:ladder} adds the information the color alone
cannot show: basis size, overlap conditioning, and the measured word universe.
The results do not support a blanket superiority statement.

\begin{table*}[t]
\caption{Selected exact-arithmetic tradeoffs.  Errors are in the indicated
Hamiltonian unit.  $W$ is the union word count of Eq.~\eqref{eq:word-universe}.
The Krylov widths are obtained from the identity of
Sec.~\ref{sec:krylov-width} rather than from the quadratic element route, at a
$10^{-8}$ coefficient threshold under which the A-CASE counts are unchanged.
The comparator is the strongest available matched or nearby-budget subspace
baseline, not necessarily a method A-CASE beats.}
\label{tab:ladder}
\scriptsize
\begin{ruledtabular}
\begin{tabular}{lccccccccc}
System & \multicolumn{4}{c}{A-CASE} & Comparator & \multicolumn{4}{c}{Comparator result} \\
 & $M$ & $|\Delta E|$ & $\kappa_S$ & $W$ & method & $M$ & $|\Delta E|$ & $\kappa_S$ & $W$ \\
\colrule
H$_4$, 0.9 \AA & 9 & $3.02\times10^{-3}$ Ha & 1 & 7371 & Krylov & 9 & $1.05\times10^{-8}$ Ha & $6.6\times10^{10}$ & 4224 \\
H$_4$, 1.8 \AA & 9 & $3.92\times10^{-2}$ Ha & 1 & 7715 & Krylov & 9 & $1.54\times10^{-5}$ Ha & $2.6\times10^{10}$ & 4224 \\
H$_2$O (4e,4o), $2R_e$ & 9 & $5.54\times10^{-2}$ Ha & 1 & 7783 & Krylov & 9 & $7.07\times10^{-3}$ Ha & $6.0\times10^{7}$ & 8192 \\
H$_2$O (8e,6o) & 9 & $1.42\times10^{-2}$ Ha & 1 & 143117 & gen.-coord. & 9 & $4.24\times10^{-2}$ Ha & 1 & 232515 \\
Hubbard $2\times2$ & 26 & $1.07\times10^{-14}$ $t$ & 1 & 15191 & Krylov & 9 & $2.64\times10^{-3}$ $t$ & $9.7\times10^{9}$ & 3968 \\
Hubbard $2\times3$ & 26 & $1.09$ $t$ & 1 & 85264 & gen.-coord. & 9 & $3.16$ $t$ & 1 & 27870 \\
Kitaev $2\times2$ & 6 & $8.12\times10^{-13}$ $K$ & $3.8\times10^{4}$ & 140 & Krylov & 9 & $6.57\times10^{-14}$ $K$ & $3.3\times10^{4}$ & 140 \\

\end{tabular}
\end{ruledtabular}
\end{table*}

On equilibrium H$_4$, A-CASE at $M=9$ improves substantially over the
same-size fixed QSE and generator-coordinate bases, but exact ADAPT-VQE is
slightly better and fixed Krylov is better by orders of magnitude.  Krylov's
overlap condition number is $6.6\times10^{10}$, versus $1$ for A-CASE.  The
same pattern is sharper for stretched H$_4$ and stretched H$_2$O (4e,4o):
Krylov buys energy accuracy with near-linear dependence, while A-CASE remains
conditioned but does not reach chemical accuracy from a determinant reference.
ADAPT-VQE's margin over A-CASE also widens on those rungs rather than
staying slight---$7.4$ against $39.2$ mHa on stretched H$_4$, and $11.9$
against $55.4$ mHa on H$_2$O---so from a determinant reference the $M=9$
A-CASE arm is the weaker of the two adaptive methods wherever both run on a
fermionic rung.

With the Krylov widths of Sec.~\ref{sec:krylov-width} in hand, the third leg
of the trade can finally be read, and it does not favor A-CASE.  On both H$_4$
rungs the Krylov arm needs $4224$ words against A-CASE's $7371$ and $7715$,
and on the Hubbard plaquette $3968$ against $15191$; on the Kitaev cluster the
two coincide at $140$, which they must, since A-CASE there \emph{is} a pruned
Krylov construction.  The single rung where the operator-generated basis is
narrower is stretched H$_2$O (4e,4o), at $7783$ against $8192$.  So at this
resolution Krylov is more accurate on every one of these rungs and narrower on
most of them.

That comparison is resolution-dependent, and Sec.~\ref{sec:matched} shows it
reversing on the equilibrium H$_4$ rung: the same retained subspace generated
at word rather than determinant resolution costs $2240$ words, below Krylov's
$4224$, at an unchanged energy and $\kappa_S=1$.  The ladder rows above are all
determinant-resolution, so they price one representation of A-CASE rather than
the method's floor, and every cross-method $W$ in this paper should be read
with its generator resolution attached.  The honest statement is that this is a
trade between energy accuracy and conditioning, with measurable support a third
axis whose comparison is not settled by these rows, and not a win by basis
count.

Against the fixed local bases the width comparison runs the other way, which
is the comparison the ledger was built to make: on H$_2$O (8e,6o) A-CASE is
both more accurate than the generator-coordinate arm and narrower, $143117$
words against $232515$.

Level-4 configuration generators change the $2\times2$ Hubbard result:
A-CASE reaches the sector ground state at $M=26$ with
$\kappa_S=1$.  The same pool adds no selected level-4 direction on
$2\times3$, leaving a $1.09t$ error at the same $M$.  Thus the $2\times2$
closure demonstrates that the missing span can be supplied, while the
$2\times3$ failure demonstrates that the current selection rule cannot
reliably discover a useful competing-order configuration.  On the Kitaev
cluster, A-CASE selects six Hamiltonian-power directions and reaches the exact
energy, but its $\kappa_S=3.8\times10^4$ resembles Krylov because in that case
it \emph{is} a pruned Krylov construction.

The finite-shot growth rows make the same warning more concrete.  Eight-qubit
certified-growth attempts spend $1.5\times10^8$ to $4.6\times10^8$ physical
shots and frequently stop or abstain with inaccurate energies.  A noisy
thresholded overlap can even yield an energy below the exact reference; this
does not violate Rayleigh--Ritz because the measured pencil no longer
represents an exact subspace projection.  Sample-split certification of a
growth coupling must not be reported as certification of the final energy.

\subsection{Whole-pipeline response uncertainty}
\label{sec:results-response}

\begin{figure}[t]
    \includegraphics[width=\columnwidth]{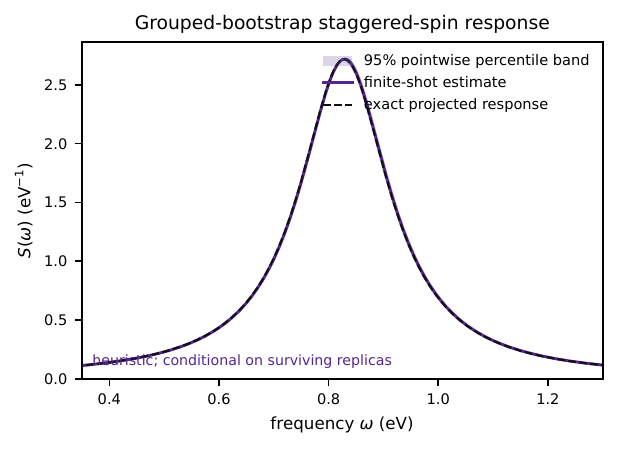}
    \caption{Lorentzian-broadened staggered-spin line with
    $\eta=0.1$ eV.  The band is pointwise across frequency, heuristic, and
    conditional on replicas that reproduce the pipeline's rank and root
    identity.}
    \label{fig:response}
\end{figure}

The response example uses the four-qubit dimer, 63 measured words, 25 QWC
groups, and 8,000 shots per group, for 200,000 physical shots.  All 200
bootstrap replicas preserve the rank and isolated ordered root in this
well-conditioned case.  Table~\ref{tab:response} and
Fig.~\ref{fig:response} compare the exact projected result with the finite-shot
point estimate and the 95\% percentile interval.

\begin{table*}[t]
\caption{Finite-shot dimer response.  The interval is a grouped-bootstrap
percentile diagnostic, not a finite-sample confidence certificate.  Acceptance
is $200/200$; the interval is conditional on the surviving replicas.}
\label{tab:response}
\begin{ruledtabular}
\begin{tabular}{lccc}
Quantity & exact & measured & percentile interval \\
\colrule
Gap $\omega$ (eV) & 0.828427125 & 0.829179080 & [0.826135816, 0.832645194] \\
Weight $w$ & 0.853553391 & 0.854210229 & [0.849206737, 0.858267589] \\
$\chi(0)$ (eV$^{-1}$) & 2.060660172 & 2.060375736 & [2.056794508, 2.062132618] \\

\end{tabular}
\end{ruledtabular}
\end{table*}

The exact values lie inside all three reported intervals in this one run.  That
observation is a consistency check, not a coverage study.

The $200/200$ acceptance also means that this run alone does not exercise the
failure accounting of Sec.~\ref{sec:bootstrap}: nothing is rejected, so the
machinery is described rather than demonstrated.  We therefore repeat it as a
controlled comparison in which conditioning is the only variable.  The system,
observable, word set, group count, shot budget, basis size, and both seeds are
held fixed; the determinant excitations are replaced by the Hamiltonian powers
$I,H,H^2,H^3$, which span the same four-dimensional sector with an overlap
condition number four orders larger.  Table~\ref{tab:conditioning} reports both.

\begin{table*}[t]
\caption{The same bootstrap at two conditionings.  Both rows use the dimer,
the staggered-spin observable, $M=4$, $63$ words, $25$ QWC groups,
$8{,}000$ shots per group, and the same seeds; only the generator family
differs.  Failure columns are the replica counts excluded for a changed
thresholded rank, a root collision, and a solver failure.  ``Width'' is the
$\chi(0)$ interval width and ``covers'' records whether the exact projected
$\chi(0)$ lies inside it.}
\label{tab:conditioning}
\begin{ruledtabular}
\begin{tabular}{lccccccc}
Generator family & $\kappa_S$ & accepted & rank & root & solver & $\chi(0)$ width & covers \\
\colrule
determinant excitations & 1.02 & 200/200 & 0 & 0 & 0 & $5.34\times10^{-3}$ & yes \\
Hamiltonian powers to order 3 & $9.3\times10^{3}$ & 139/200 & 61 & 0 & 0 & $8.18\times10^{-2}$ & yes \\

\end{tabular}
\end{ruledtabular}
\end{table*}

\begin{figure}[t]
    \includegraphics[width=\columnwidth]{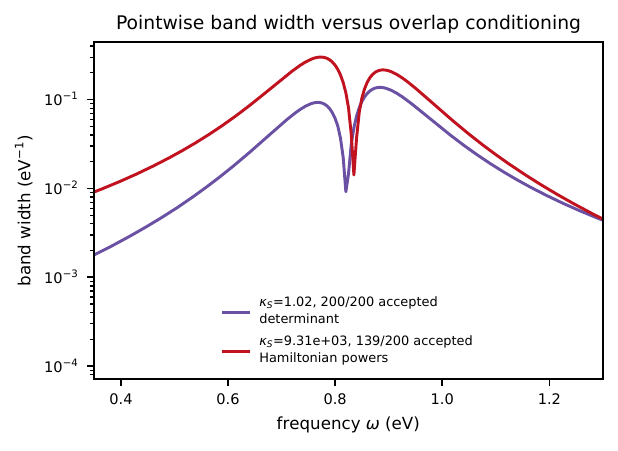}
    \caption{Pointwise width of the $95\%$ band across frequency for the two
    rows of Table~\ref{tab:conditioning}.  The absolute spectra are visually
    indistinguishable, so the width is plotted directly.  The cusp sits at the
    line center, where an uncertain gap moves the Lorentzian along the axis
    without changing its height.}
    \label{fig:conditioning}
\end{figure}

The ill-conditioned arm rejects $61$ of $200$ replicas, all of them for a moved
thresholded rank, and its surviving-replica $\chi(0)$ interval is roughly
fifteen times wider.  Figure~\ref{fig:conditioning} shows the same separation
holding across the whole frequency axis rather than at a single summary point.
Two things follow.  The diagnostic does respond to
conditioning rather than reporting a comfortable width regardless, which is the
minimum one should demand of it.  But the reported width is still conditional
on the $139$ replicas that reproduced the pipeline's rank; the $61$
excluded draws are precisely the ones a rank-stable summary cannot represent,
and the conditional interval does not bound what they would contribute.  This is why the
acceptance rate belongs beside every interval, and why larger systems must
report the rank, root-collision, and solver-failure counts before the width of
a surviving-replica interval is interpreted.

\section{Discussion}
\label{sec:discussion}

\subsection{What is established}
\label{sec:established}

Three claims are directly supported.  First, a real active-space interchange
record can be parsed, mapped, and checked against independent determinant FCI
without PySCF or OpenFermion at solver runtime.  Second, one reference state
and one word cache are sufficient to construct the projected energy,
observables, and response matrices; state-specific circuits and materialized
Ritz statevectors are not intrinsic to this operator-generated formulation.
Third, adaptive selection can preserve good overlap conditioning while
improving materially over fixed local QSE and generator-coordinate bases on
some rungs.

A fourth claim is supported by Table~\ref{tab:warm}: for the frozen H$_4$
instance, determinant-excitation family, and nine-vector A-CASE budget, a
two-rotor ADAPT reference is sufficient to cross chemical accuracy.  This is
also the one place where the two adaptive methods compose rather than compete,
since the ADAPT state is a worse energy answer on its own than the A-CASE run
it improves.  It is not a total-resource advantage claim.

The results also establish useful negative boundaries.  At the matched $M=9$
budget, fixed Krylov is the energy winner on every fermionic rung where it was
configured, despite severe conditioning; the exception is the $2\times2$
Hubbard cluster, where level-4 A-CASE at $M=26$ reaches the sector ground state
that Krylov misses by $2.6\times10^{-3}t$.  Krylov also needs fewer measured
words than determinant-resolution A-CASE on most rungs where both counts are
available, though Sec.~\ref{sec:matched} shows that ordering reversing on
equilibrium H$_4$ once the same subspace is generated at word resolution, so
the width comparison is a statement about a representation and not about the
method.  A second boundary belongs beside it: the measurement cost the ledger
reports is the retained subspace's, and scoring the candidates that were
rejected on the way to it cost $15783$ words against the $7371$ quoted.  From a
determinant reference the default A-CASE pool does not reach chemical accuracy
on H$_4$ or H$_2$O at $M=9$.  Compound/configuration generators solve the
$2\times2$ Hubbard case but do not transfer to $2\times3$.  Finite-shot
response bands diagnose the implemented nonlinear pipeline but do not carry a
finite-sample coverage theorem, and Table~\ref{tab:conditioning} shows their
acceptance rate falling to $139/200$ as soon as the overlap is ill-conditioned.
The adaptive-GEP architecture itself is not claimed as new: ADAPT-GCIM is the
closest precedent.  The exact matched-pool comparison above implements its
fixed-angle selector and $M=2k$ generating-function basis, while keeping its
off-diagonal $H/S$ measurement burden separate from the fixed-reference bank.
It is not a reproduction of the original paper's full benchmark setting.

\subsection{What remains open}
\label{sec:open}

The principal algorithmic question is whether candidate families can reduce
$M$ \emph{and} $W$ while keeping $\kappa_S$ controlled on active spaces beyond
eight qubits.  A promising direction is block or state-averaged growth against
several Ritz roots, because a ground-state-selected basis is not automatically
response-complete.  A second direction is to replace the candidate-by-candidate
two-dimensional score with a residual block or a preconditioned correction,
closer to Davidson methods, while preserving the shared-word accounting
\cite{tkachenko2024davidson}.

Statistically, three layers must remain separate.  Regularizing a measured
$S$ can stabilize a solve, but does not by itself restore a variational bound.
A bootstrap can reveal nonlinear spread, but survivor conditioning can hide
the worst draws.  A genuine response certificate needs simultaneous control
of a correlated matrix pencil, a root-isolation event, and a nonlinear
observable map.  Existing QSD perturbation and sampling analyses provide the
right starting point \cite{epperly2022theory,lee2024sampling}, but the complete
construction is not supplied here.

Finally, a materials claim requires a real upstream calculation and comparison
with an accepted correlated solver on the same downfolded Hamiltonian.  The
current effective-Hamiltonian JSON proves the software boundary, and FCIDUMP
proves a real interchange boundary.  Neither substitutes for validating the
downfolding, double-counting correction, interaction model, or observable
embedding for a material.

\section{Reproducibility and evidence ledger}
\label{sec:reproducibility}

The implementation, frozen inputs, raw benchmark records, table and figure
generators, tests, and this manuscript are maintained together in the
private development repository \texttt{clifford\_qc} while its operator,
measurement, subspace, and backend interfaces are being stabilized.  Public
access to the moving development branch is not claimed.  For independent
review, the authors will provide editors and referees, on request, an
access-controlled frozen snapshot that identifies the exact source revision
and contains the code, data, environment specification, and artifact
generators used here.  A tagged archival release with a persistent identifier
is intended once the framework interfaces and evidence contracts are stable.
Thus private development changes the distribution channel, not the executable
evidence standard described below.

The H$_4$ record contains the FCIDUMP SHA-256 digest and independent PySCF
version and energies,
and the warm-start record repeats them for every reference state in
Table~\ref{tab:warm} together with the ADAPT operator labels.  Both finite-shot
response records contain both random seeds, the shots per group, the QWC group
count, the requested and accepted replicas, every failure category, and the
full plotted frequency arrays.  The Krylov width record carries the whole
threshold sweep, not only the value quoted in Table~\ref{tab:ladder}.
The matched record stores both ADAPT-GCIM matching rules, the selected
excitation sequence, fixed angle, nominal and retained ranks, every trajectory
energy, basis-circuit rotor depths, and separate Hamiltonian and overlap pair
counts.

The independent checks are: (i) the dimer closed forms in
Table~\ref{tab:dimer}; (ii) determinant FCI versus the mapped H$_4$ sector;
(iii) OpenFermion versus the direct chemist-notation mapping; (iv) exact
infinite-shot reconstruction through the finite-shot coefficient maps;
(v) replica accounting, for which every requested draw must either succeed or
enter exactly one failure category; and (vi) the Krylov width identity of
Eq.~\eqref{eq:krylov-collapse} against direct element enumeration, which also
reproduces the ladder's own tracked counts on the two rungs where that route
finishes.  The repository test suite and
\texttt{python -m clifford\_qc.verify} exercise algebraic, fermionic,
subspace, effective-model, FCIDUMP, and response invariants.

For clarity, the claim ledger is:
\begin{itemize}
 \item \emph{exact arithmetic}: dimer identities, FCIDUMP mapping, sector
 energies, exact subspace energies, and exact resource counts;
 \item \emph{finite sample}: only declared sample-split candidate-growth
 decisions, conditional on their assumptions;
 \item \emph{heuristic}: bootstrap response intervals;
 \item \emph{not claimed}: novelty of adaptive GEPs, superiority to
 ADAPT-GCIM, DFT replacement, material prediction, scaling advantage,
 hardware demonstration, or quantum speedup.
\end{itemize}

\section{Conclusion}
\label{sec:conclusion}

A-CASE provides a narrow but complete downstream workflow: an effective
many-body Hamiltonian enters through a versioned or FCIDUMP boundary; one
reference state supplies a cached operator-generated subspace; the subspace
returns energies, correlations, and response; and finite-shot response is
accompanied by an explicit evidence label and failure accounting.  The
independent H$_4$ mapping check and closed-form dimer oracle make the workflow
more than a self-consistency demonstration.

The benchmark outcome is intentionally nonuniform.  A-CASE improves on some
fixed local subspaces while fixed Krylov remains more accurate but often
ill-conditioned; a configuration extension closes one Hubbard cluster and
fails on the next.  This mixed result is the useful one.  It identifies
conditioning, measurement width, and response completeness as coequal design
objectives, and it prevents a small basis or a narrow bootstrap band from
being mistaken for evidence that the full correlated problem has been solved.
We make no quantum advantage claim.

\bibliography{references}

\end{document}